# Oxophilic Silver-Based Nanoparticles with Low Pd-Au Loading for Ethanol and Glycerol Electrooxidation in Alkaline Media

*Tuani Carla Gentil,[*] Camilo Andrea Angelucci, Bruno Lemos Batista, Camila Neves Lange, Handro S. N. Lourenço, Mauro Coelho dos Santos, Vinicius Del Colle,[*] and Germano Tremiliosi-Filho[*]*

T.C. Gentil, H. S. N. Lourenço, G. Tremiliosi-Filho

Chemistry Institute of São Carlos, University of São Paulo

Avenida Trabalhador São-carlense, 400, São Carlos, SP, Brazil

E-mail: tuanigentil@usp.br; germano@iqsc.usp.br

V. Del Colle

Aeronautics Technological Institute, Chemistry Department

Praça Marechal Eduardo Gomes, 50 São José dos Campos, São Paulo, Brazil

E-mail: delcolle@ita.br

C.A. Angelucci, B.L. Batista, C.N. Lange, M.C. Santos

Federal University of ABC, Center for Natural and Human Sciences

Av. dos Estados, 5001, 09210-580, Santo André, São Paulo, Brazil

The electrocatalytic activity of oxophilic Ag nanoparticles, combined with small amounts of Pd and Au, was investigated for ethanol oxidation reactions (EOR) and glycerol oxidation reactions (GOR) in alkaline media. The EOR and GOR results revealed competitive current densities and less positive onset potentials for the AgPd/C and AgPdAu/C electrocatalysts—both containing 5 wt% Pd—compared to the commercial Pd/C catalyst, which has a significantly higher loading of the costly noble metal (20 wt%). In situ FTIR analyses during EOR confirmed that ethanol is initially adsorbed as acetylated species ($CH_3CO_{ads}$), which are subsequently oxidized to acetate ions, the main stable product in alkaline medium. However, the incorporation of Pd and Au into the Ag matrix did not significantly alter the reaction mechanism. During GOR, the in situ FTIR studies demonstrated that catalyst composition influences the oxidation pathways: Pd-rich surfaces favor oxalate formation, while a significant presence of Ag promotes deeper oxidation (up to carbonate), with the AgPdAu ternary catalyst exhibiting intermediate behavior. One key benefit is Ag's lower susceptibility to irreversible adsorption of reaction byproducts, which enhances electrocatalyst durability. Thus, surface segregation of Ag at high potentials can modify the catalytic surface reactivity, affecting both stability and efficiency.

# 1. Introduction

Renewable energy sources are at the heart of the transition to more carbon-efficient and sustainable energy systems.[1–3] According to the International Energy Agency (IEA), approximately 3,700 GW of new renewable capacity is expected to become operational between 2023 and 2028, driven by supportive policies in more than 130 countries, with significant milestones projected in the renewable energy landscape over the next five years.[4]

In this context, direct fuel cells offer a viable alternative, delivering high efficiency and low pollutant emissions, convenient storage, simple refuelling, and high energy density [5] in addition to enabling the direct use of biofuels, e.g., ethanol[5,6] and glycerol,[7–9] becoming an attractive opportunity for sustainable energy consumption and storage.

Glycerol is a byproduct of industrial biodiesel production, generated in large quantities, and this surplus presents an opportunity for its use. In Brazil, for instance, approximately 80,000 tons of glycerol are produced annually, while consumption ranges from 30,000 to 40,000 tons per year. As a result, glycerol is widely available and inexpensive, which justifies further studies on its potential use as a fuel.[9–11]

The significance of the ethanol oxidation reaction (EOR) in the context of hydrogen energy has driven extensive research. Enhancing EOR kinetics is directly related to catalyst performance, as effective catalysts accelerate reaction rates, mitigate polarization potential, and improve the overall efficiency of water electrolysis/fuel cells.[12]

Both ethanol and glycerol play an important role in shaping Brazil's energy mix. Ethanol is particularly significant, as the country is recognized as a global leader in the production, distribution, and utilization of sugarcane-based ethanol and glycerol, a byproduct of biodiesel production, which has also been attracting growing interest due to its potential applications in the energy and chemical industries.[13]

The development of cost-effective and efficient electrocatalysts for the oxidation of small organic molecules, such as ethanol and glycerol, is crucial for advancing sustainable energy technologies. Noble metals, such as palladium (Pd), platinum (Pt), and gold (Au), exhibit excellent catalytic properties but are limited by their high cost and scarcity.[14] Recent research has focused on minimizing the use of these metals by incorporating them into metal matrices, particularly silver (Ag), to create bimetallic or trimetallic nanoparticles with enhanced catalytic performance.[15,16]

*Silver was selected as the primary matrix due to its relatively low cost, high oxophilicity, and electrochemical stability in alkaline media. Although Ag is not intrinsically*

*active toward ethanol electrooxidation, it exhibits measurable activity toward glycerol oxidation and, more importantly, promotes the formation of surface hydroxylated species (Ag–OH/Ag$_2$O), which assist alcohol oxidation on adjacent noble-metal sites* [17]. *Gold was incorporated in small amounts (5 wt%) to probe its electronic and oxophilic effects, while palladium remains the main high-cost metal in this system* [18]. *In AgPd/C and AgPdAu/C, Pd loading is only 5 wt%, significantly lower than in the reference commercial standard Pd/C (20 wt%).*

Studies have demonstrated that incorporating small amounts of noble metals into Ag-based nanoparticles can significantly enhance their electrocatalytic activity. For instance, PdAg nanoparticles with varying Pd/Ag ratios have been synthesized, showing that a Pd/Ag molar ratio of 4:1 yields the highest catalytic current density for ethanol oxidation, attributed to the synergistic effects between Pd and Ag, as well as the reduced particle size. Similarly, AgPt nanoalloys with a 1:3 Ag/Pt ratio have exhibited superior catalytic activity and durability for ethanol oxidation, highlighting the beneficial role of Ag in enhancing the performance of Pt-based catalysts [19,20]. In the context of glycerol oxidation, Ag nanoparticles modified with low amounts of Pt (as low as 0.5 wt%) have achieved significant activity.[15]

Noble metals, such as platinum (Pt) and palladium (Pd), exhibit high activity in catalyzing biofuel oxidation reactions; however, their high cost and susceptibility to catalytic poisoning remain significant obstacles.[5] An effective alternative in this context is the incorporation of oxophilic metals, e.g., silver (Ag),[15,17,20,21] gold (Au),[21–23] nickel (Ni),[24,25] rhodium (Rh),[26,27] ruthenium (Ru),[28] and tin (Sn),[6,29,30] into the electrocatalyst composition. In alkaline media, this approach promotes the formation of OH radicals, enhances the activity for biofuel oxidation, and minimizes the production of byproducts such as CO. Furthermore, alloys with oxophilic metals, such as Ag or Au, improve the CO tolerance and electrocatalytic activity of the materials due to synergistic and electronic change mechanisms.[17,31,32]

As mentioned above, Pd is an excellent electrocatalyst for alcohols in an alkaline medium. The presence of silver in Pd-based electrocatalysts can further increase the surface reactivity of molecules by modifying the d-band center of Pd, resulting from the larger lattice parameter of Ag (a = 4,09 Å).[25] In addition to Pd, Au has also demonstrated notable activity for the oxidation of alcohols in an alkaline medium, which can be attributed to the minimal formation and adsorption of toxic species on its surface.[23] Additionally, alkaline medium offers advantages such as faster oxygen reduction reaction kinetics, reduced electrode

material corrosion, and enhanced stability, allowing the use of a broader range of electrode materials.[3,33]

The improved performance of these bimetallic systems is often attributed to electronic and geometric effects. The presence of Ag can modify the electronic structure of the noble metal, facilitating the adsorption and activation of reactant molecules. Additionally, the formation of alloyed or core-shell structures can influence the distribution of active sites and the stability of the catalysts. For example, PdAu bimetallic catalysts with controlled geometries have demonstrated enhanced ethanol oxidation activity compared to commercial Pt/C catalysts, highlighting the importance of structural design in determining catalyst performance.[34]

High-surface-area carbonaceous materials contribute to the electrical conductivity of the catalyst and the homogeneous dispersion of metal nanoparticles. However, carbon undergoes dissolution and degradation due to the influence of electrolytes in both acidic and alkaline environments.[35] In alkaline electrolyte, OH$^-$ ions can interact with defects in the carbon structure, leading to its fragmentation, which may result in the loss of electrical contact between the metal nanoparticles and the electrode, as well as the aggregation or leaching of the metal nanoparticles from the catalyst, thereby decreasing the active surface area and reducing catalytic activity.[35,36]

Despite these advancements, challenges remain in optimizing the composition, structure, and synthesis methods of low-content precious metal catalysts. Further studies are needed to understand the long-term stability of these materials under operational conditions and to explore the scalability of their synthesis. Moreover, *in situ* characterization techniques, such as Fourier-transform infrared spectroscopy (FTIR), can provide valuable information about intermediate species and reaction processes, guiding the rational development of more effective catalysts. Therefore, this research aims to synthesize simple, binary, and ternary nanoparticle electrocatalysts comprising Ag, Pd, and Au, supported on Vulcan XC-72 carbon, with minimal amounts of Pd and Au, characterizing their structural and electronic properties, and evaluating their electrocatalytic performance toward the oxidation of glycerol and ethanol.

## 2. Results and Discussion

### 2.1. Characterizations of Ag–Pd–Au Nanoparticles Electrocatalysts

To investigate the structural features of the synthesized electrocatalysts— Ag/C, AgPd/C, AgAu/C, and AgPdAu/C—X-ray diffraction (XRD) analyses were conducted

(Figure 1), with detailed individual XRD patterns of each material shown in the supplementary material (Figures S1 – S4). The patterns were indexed by comparison with reference data from the database, the Joint Committee on Powder Diffraction Standards (JCPDS).

A broad peak at 2θ ≈ 25°, as consistently observed across all samples, corresponds to the {002} plane of the hexagonal structure of the Vulcan XC-72 carbon support.[37,38] Superimposed upon this background, the metallic phases exhibit distinct diffraction peaks characteristic of face-centered cubic (fcc) silver[31,39] located, respectively, at 2θ = 38.05°, 44.20°, 64.43°, 77.36°, and 81.51°. Given that silver is the primary metallic component in all electrocatalysts, these features dominate the XRD profiles.

The incorporation of secondary metals, such as palladium and gold, into the Ag matrix was initially assessed by examining subtle shifts in XRD peak positions. However, due to the similar lattice parameters of Ag and Au, the identification of Au incorporation based solely on lattice parameter variations is limited. Additionally, the database indicates that the diffraction pattern of Au (JCPDS Nº. 65-2870) shows peaks that almost completely overlap with those of Ag. As a result, to validate the incorporation and distribution of Au in the electrocatalysts, complementary physicochemical characterization techniques were performed and will be discussed later.

Moreover, the low concentrations of Pd and Au, which have the FCC crystal structure, coupled with the instrumental parameters employed during data acquisition (step size of 0.0200°/s and scan rate of 2°/min), restrict the resolution necessary to discern subtle peak shifts or splitting in the primary diffraction reflections. However, a detailed analysis of the AgPd/C diffractogram reveals asymmetric broadening of the (111) peak, indicative of a structurally heterogeneous phase likely influenced by the partial incorporation of Pd into the Ag lattice.[40]

To further investigate this phenomenon in greater detail, additional higher-resolution X-ray diffraction measurements were performed within a narrowed 2θ range of 36° to 42°, using a finer step size of 0.005° s$^{-1}$ and a scan of 0.05° min$^{-1}$ (Figures S5-S6). The resulting deconvolution revealed two distinct peaks near 2θ = 40°: for AgPd/C, a primary peak centered at 38.10° and a secondary, less intense peak at 38.75°; for AgPdAu/C, a primary peak at 38.20° and a secondary, also less intense peak at 39.14°. This dual-peak structure supports the presence of two compositionally distinct (111) domains, indicating that Pd *is present in compositional heterogeneity* distributed within the Ag matrix. Instead, the data suggest the formation of Pd-rich regions or segregated phases that were not fully incorporated into a uniform solid solution, which is consistent with the well-known complete miscibility of the Au–Ag binary system, as predicted by the Hume–Rothery criteria.[40] This initial evidence of incomplete alloying will be further investigated through electrochemical characterization, as discussed in subsequent sections. *At very low Au loading (5 wt%), Au may be below the detection limit of XRD as a separate phase while still appearing locally segregated at the nanoscale in TEM-EDS. From XRD, it was possible to estimate the interplanar distance and lattice parameter using Bragg's law* [14,25] *taking into account the (111) peak, as can be seen in Figure S7 and Table S1. The small shifts observed for values greater than 2θ are consistent with limited incorporation of Pd and Au at low concentrations.*

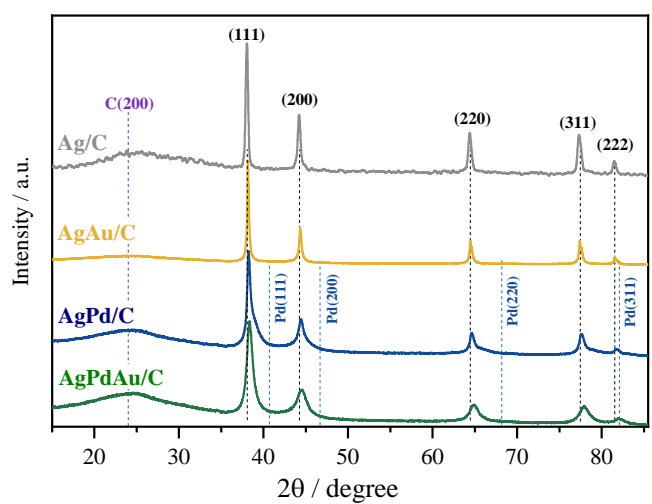

**Figure 1.** XRD patterns of Ag/C; AgPd/C; AgAu/C, and AgPdAu/C. Vertical dashed lines indicate the standard 2θ positions for the characteristic reflections of bulk Ag and Pd, based on JCPDS reference data.

The elemental composition of the synthesized electrocatalysts was determined through a combination of energy-dispersive X-ray spectroscopy (EDS) and inductively coupled plasma mass spectrometry (ICP-MS), as summarized in Table 1. The semi-quantitative EDS results for Pd were in good agreement with the quantitative values obtained by ICP-MS. According to the ICP-MS results, the mass percentages of the high-cost metals (Pd and Au) closely approximated their respective nominal loadings, confirming the effectiveness of the synthesis protocol in controlling the targeted composition. However, in the case of Ag and Au, the EDS-derived mass percentages exhibited greater deviation from the expected stoichiometry, which may be attributed to local inhomogeneities within the catalyst matrix or surface-enriched regions that affect the spatial resolution of EDS measurements. This observation is supported by transmission electron microscopy (TEM) analysis of the AgAu/C electrocatalyst (Figure 3c), which reveals a non-uniform dispersion of nanoparticles across the carbon support.

A complementary study was conducted to evaluate the total metal content and metal loading on carbon support using thermogravimetric analysis (TGA). The TGA technique allows the evaluation of mass loss as a function of temperature under controlled atmospheric conditions.

**Table 1.** Elemental composition of electrocatalysts estimated by EDS and ICP-MS.

| Electrocatalyst | EDS (%) | | | ICP-MS (%) | | Nominal mass ratio (%) | | | |
|---|---|---|---|---|---|---|---|---|---|
| | Pd | Ag | Au | Pd | Au | C | Pd | Ag | Au |
| AgPd/C | 7.4 ± 5.7 | 22.0 ± 4.2 | (-) | 6.3 ± 0.01 | (-) | 55 | 5 | 40 | (-) |
| AgAu/C | (-) | 21.1 ± 5.5 | 1.7±0.6 | (-) | 4.5 ± 0.2 | 55 | (-) | 40 | 5 |
| AgPdAu/C | 2.4 ± 1.0 | 15.9 ± 6.9 | 1.1±1.1 | 6.7 ± 0.1 | 4.6 ± 0.2 | 55 | 5 | 35 | 5 |

Figure 2 shows TGA curves of the synthesized electrocatalysts and the reference material. A mass loss was observed at the beginning of the analysis at low temperatures, which may be related to the moisture present in the samples.[25,41] Amorphous carbon presents degradation at lower temperatures (~500 °C), as evidenced by a considerable mass loss.[41,42] The metal and carbon composition estimated by TGA is close to the nominal mass ratio (Table 2).

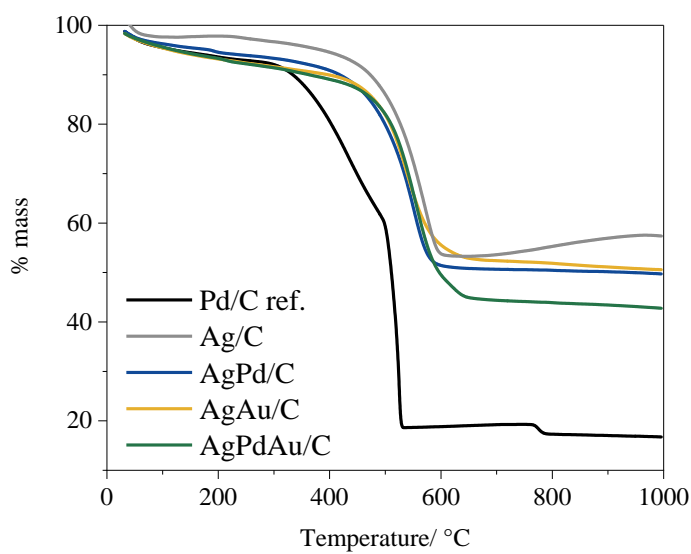

**Figure 2.** TGA curves of the synthesized and reference electrocatalysts.

**Table 2.** Metal and carbon composition estimated by TGA.

| Electrocatalyst | % metal | % C -Vulcan XC-72 |
|:---:|:---:|:---:|
| Pd/C ref. | 18.4 | 81.6 |
| Ag/C | 53.5 | 46.5 |
| AgPd/C | 49.7 | 50.3 |
| AgAu/C | 50.6 | 49.4 |
| AgPdAu/C | 42.8 | 57.2 |

Figure 3 presents representative TEM micrographs, while Figure 4 presents the corresponding particle size distribution histograms of **(a)** Ag/C, **(b)** AgPd/C, **(c)** AgAu/C, and **(d)** AgPdAu/C electrocatalysts. According to the TEM images, all the electrocatalysts consist of quasi-spherical nanoparticles dispersed on the surface of the Vulcan XC-72 carbon support. Notably, the AgAu/C sample shows a higher degree of particle agglomeration compared to the other compositions. *Au(III) has a higher reduction potential than Ag(I), favouring early nucleation of Au-rich seeds that subsequently promote coalescence, especially in the absence of strong capping agents. This leads to larger aggregated particles and broader distributions in Au-containing samples. In the ternary AgPdAu system, the simultaneous reduction of three metals and possible galvanic interactions create multiple nucleation and growth events, generating a broader size distribution than in AgPd/C, where Pd addition seems to favor more uniform nucleation and smaller particles.*

For bimetallic and trimetallic systems, EDS spectra were obtained from the corresponding TEM regions to verify the composition of the nanoparticles. Although the individual metal components within each nanoparticle could not be distinctly resolved, the spectra showed the presence of the elements in the materials studied. Such distribution is better visualized through SEM-EDS elemental mapping, as displayed in Figures S8-S11, which suggests that the nanoparticles are of mixed metallic composition, consisting of silver and palladium in AgPd/C, silver and gold in AgAu/C, and silver, palladium, and gold in AgPdAu/C. *The EDS spectra for the bimetallic and trimetallic materials are shown in Figures S12-14, confirming the presence of the constituent metals of the electrocatalysts. HRTEM images of bimetallic and trimetallic materials are shown in Figure S15. The highlighted figures are magnifications of specific selected nanoparticles, used to extract their respective interplane distances. These figures reveal details of the digital magnification without increasing the microscope magnification. Furthermore, Figure S16 shows the EDX/elemental mapping using the TEM image of the AgPdAu/C material.*

The average size of the Ag/C particles is 11.5 ± 6.4 nm, with a significant presence of particles with larger particle sizes. The presence of Au in the bimetallic and trimetallic systems shifts and broadens the particle size distribution, with larger sizes displaying values of 14.8 ± 4.9 nm and 9.7 ± 8.0 nm for AgAu/C and AgPdAu/C, respectively. However, the homogeneity of particle size is more pronounced for the AgPd catalyst, presenting a narrow size distribution of the nanoparticles (AgPd/C; 8.1 ± 3.2 nm).

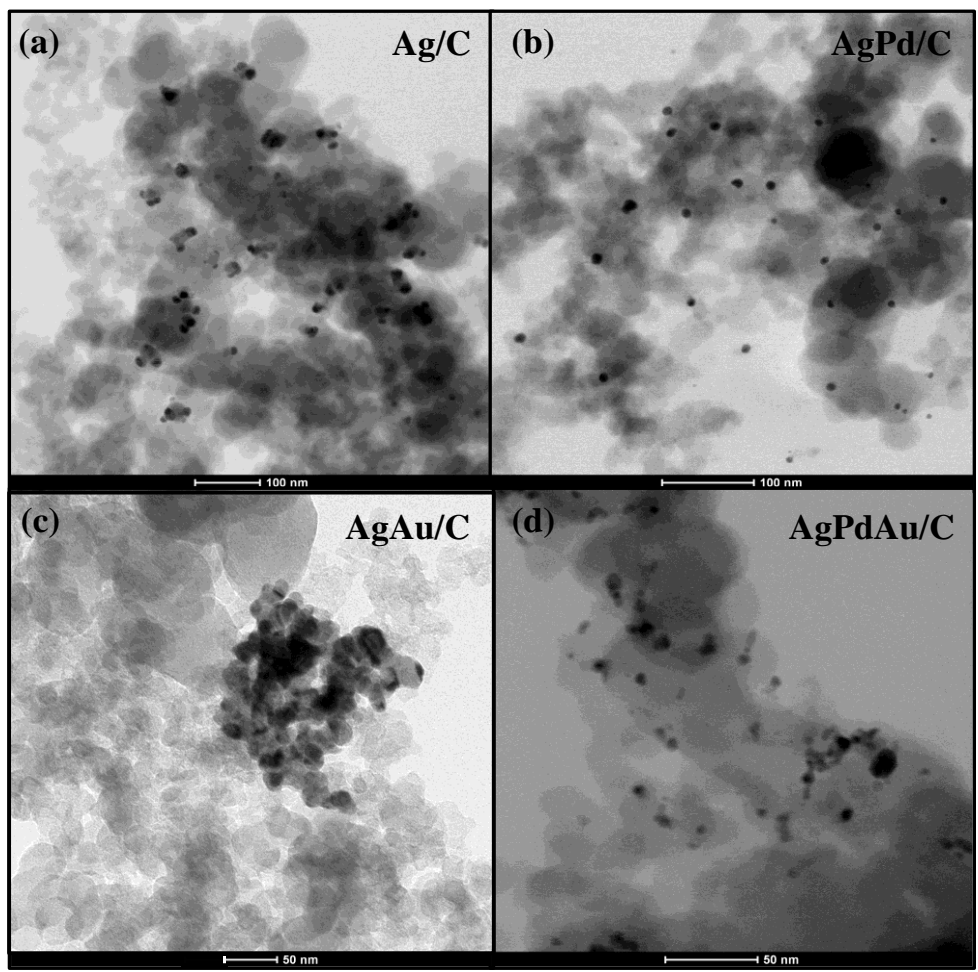

**Figure 3.** TEM images of (a) Ag/C; (b) AgPd/C; (c) AgAu/C, and (d) AgPdAu/C.

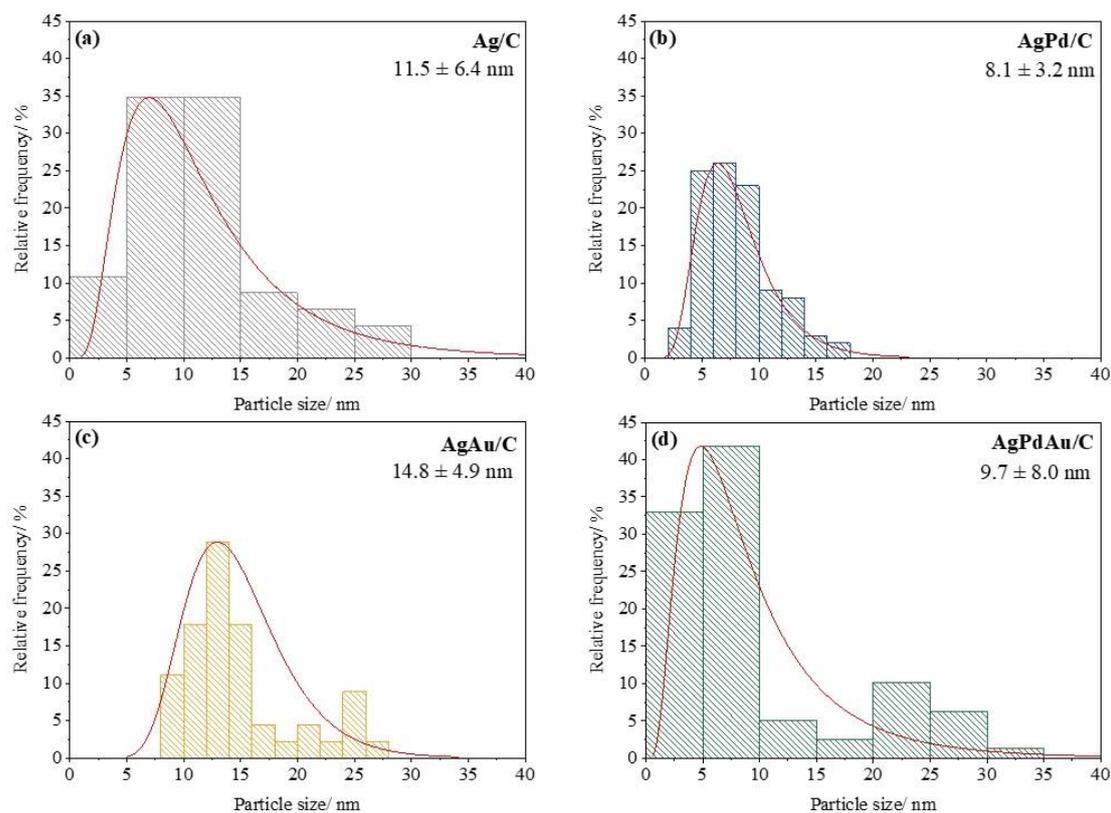

**Figure 4.** Histograms of particle sizes of the electrocatalysts (a) Ag/C; (b) AgPd/C; (c) AgAu/C, and (d) AgPdAu/C.

## 2.2. Electrocatalytic Performance Tests
### 2.2.1. Electrochemical Characterization

Figure 5 shows the blank cyclic voltammetry (CVs) profiles of the (a) Pd/C, AgPd/C, AgAu/C, and AgPdAu/C, and (b) Ag/C catalysts in KOH solution. The recorded CV profiles of Pd/C and Ag/C catalysts are characteristic of Ag and Pd metal nanoparticles supported on carbon substrates, consistent with previously reported studies.[6,21]

The CV of the Ag/C catalyst (Figure 5b) displays a broad capacitive region spanning from approximately 0 to 1.0 V, which is typical for the Ag electrode in alkaline medium.[43] A pronounced peak at 1.25 V in the forward scan is attributed to silver oxide formation, followed by *a reduction peak around 1.10 V* on the reverse scan, confirming the reversible reduction of silver oxide species. The oxidation peak (P1) at 1.17 V is attributed to the formation of a few monolayers of AgOH and $Ag^+$ species. The peaks P2 (1.25 V) and P3 (1.30 V) are attributed to the formation of a denser and more compact internal hydrated oxide and an external oxide layer, respectively.[25]

For catalysts containing Au, specifically the AgAu/C and AgPdAu/C samples, the CV profiles do not clearly distinguish Au oxide formation due to the overlap between the potential windows of Au- and Ag-oxide formation. This overlap makes it challenging to differentiate clearly between alloyed and individual metal oxidation processes solely based on CV analysis, necessitating the use of additional characterization techniques to resolve alloy compositions and surface dynamics.

Regarding Pd-containing catalysts, notably AgPd/C and AgPdAu/C, the characteristic features associated with hydrogen absorption/desorption typically occurring between 0.05 V and 0.25 V are notably absent, a result differing significantly from pure Pd-based catalysts.[6] Furthermore, a reduction peak associated with palladium oxide species is observed at approximately 0.65 V, slightly shifted toward lower potentials compared to pure Pd/C. This shift indicates a modification in the electrochemical properties of palladium, likely caused by electronic interactions with the Ag matrix, in agreement with structural analysis via XRD.

Additionally, an increase in oxidation current was observed within the potential range associated with Ag-oxide formation as the Pd content increased. This enhancement suggests dual contributions from direct silver oxidation and indirect oxidation involving Pd segregated onto the Ag surface, emphasizing the critical influence of alloy composition and structure on electrochemical behavior.[40]

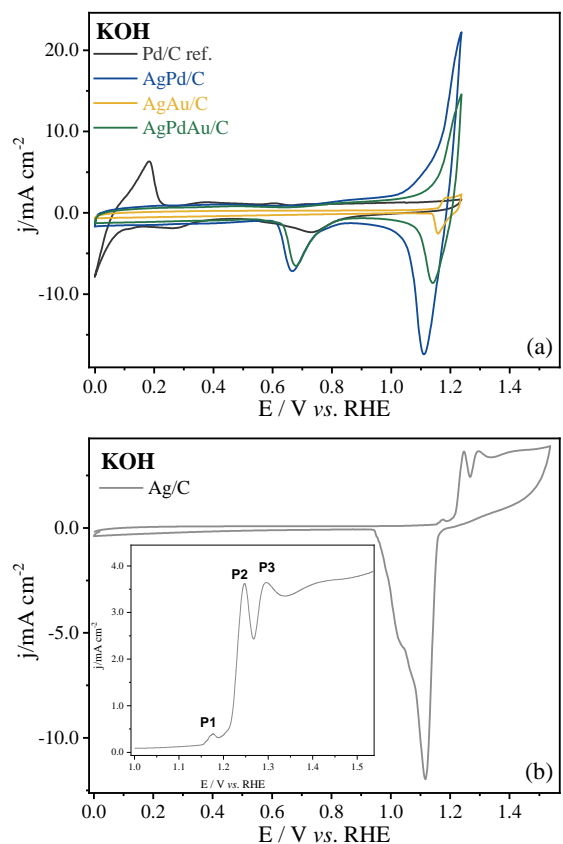

**Figure 5.** Cyclic voltammograms recorded in KOH (1.0 mol L$^{-1}$) for (**a**) Pd/C ref., AgPd/C; AgAu/C, and AgPdAu/C; and (**b**) Ag/C. ν = 20 mV s$^{-1}$

**2.2.2. Electrocatalytic Performance for EOR and GOR**

The electrocatalytic performance of the synthesized materials for EOR and GOR was investigated in 1.0 mol L$^{-1}$ ethanol in 1.0 mol L$^{-1}$ KOH, and 1.0 mol L$^{-1}$ glycerol in 1.0 mol L$^{-1}$ KOH, respectively. The activity of the electrocatalysts was compared to that of commercial Pd/C.[9] Figure 6a presents representative CVs for EOR obtained with different catalysts, where characteristic ethanol oxidation peaks were identified in the forward and reverse scans, except for AgAu/C, which exhibited no activity for EOR in an alkaline medium.[5,44] Figure 6b presents representative CVs for GOR, where characteristic glycerol oxidation peaks were also identified. In this case, AgAu/C exhibited notable activity for GOR.[45] However, a significantly shifted onset potential towards more positive values was obtained under potentiodynamic conditions compared to the others.

According to Obradović *et al.*[44] EOR studies found that the presence of Ni in the composition of Pd-based materials promotes the adsorption of OH species at lower potentials, which influences the EOR onset potential. Additionally, it facilitates the removal of reaction intermediates, regenerates the Pd active sites, and thereby accelerates the further oxidation of ethanol molecules. In this context, the presence of Ag as a co-catalyst may have an influence, shifting the onset potential for both EOR and GOR to more negative values, which would facilitate the oxidation of alcohol molecules.

Oxophilic species (such as $Ag_2O$ or $AgOH$) can increase the availability of adsorbed OH groups, which favors the breaking of the C—C bond, facilitating EOR. The presence of silver oxides can reduce CO poisoning, improving catalyst stability.[46] The presence of oxophilic species of Ag and Au nanoparticles can also favor the adsorption and activation of oxygenated species ($OH^-$), promoting the efficient oxidation of ethanol in an alkaline medium, reducing the accumulation of inactive intermediates, which results in an improvement in the reaction kinetics and increases the stability of the electrocatalysts.[46,47]

Due to the significant Ag content in the composition of the electrocatalysts, a peak was observed in the reverse scan at 1.1 V in both the EOR and GOR studies, corroborating the electrochemical characterization. This peak can be attributed to the reduction of Ag oxy-hydroxides formed during the forward scan, except for AgAu/C in the GOR, where the potential coincided with the oxidation peak potential of glycerol and its intermediates, thereby suppressing the appearance of the Ag oxy-hydroxide peak.[25]

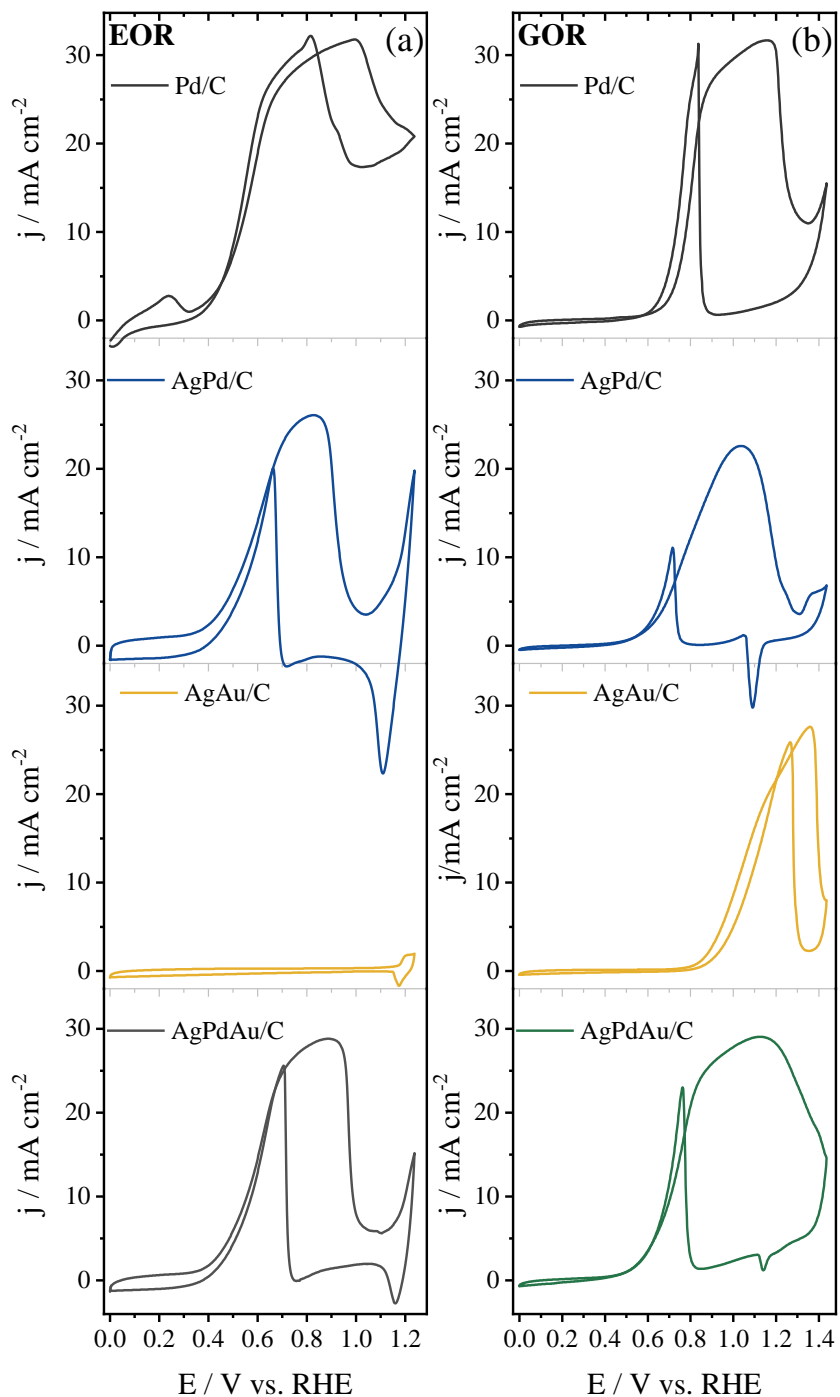

**Figure 6.** Cyclic voltammograms recorded for Pd/C ref., AgPd/C; AgAu/C, and AgPdAu/C. **(a)** ethanol (1.0 mol L$^{-1}$) and **(b)** glycerol (1.0 mol L$^{-1}$), both in KOH (1.0 mol L$^{-1}$). ν = 20 mV s$^{-1}$

The performance of the Ag/C electrocatalyst for EOR and GOR in an alkaline medium was evaluated (Figure 7) and compared to the characterization profile in 1 mol L$^{-1}$ KOH. *In the EOR studies, a oxidation peak is observed in the forward scan around 0.75 – 0.95 V for the AgPd/C and AgPdAu/C, and round 0.95 – 1.05 V for the Pd/C, followed by a peak at ca. 0.7 – 0.8 V in the reverse scan. The latter can be attributed to the oxidative desorption of*

*species formed during the forward-scan oxidation of ethanol and its intermediates* [14]. Despite the relatively low current densities, the GOR experiments using Ag/C exhibited oxidation peaks in both the forward and reverse scans, even without the use of high-cost metals such as Pd and Au. These results support the GOR studies shown in Figure 6b, where the presence of Ag in the AgPd/C, AgAu/C, and AgPdAu/C catalysts may have contributed to the observed current densities, which were very close to those of commercial Pd/C ref, despite a significant reduction in the loading of high-cost metals.

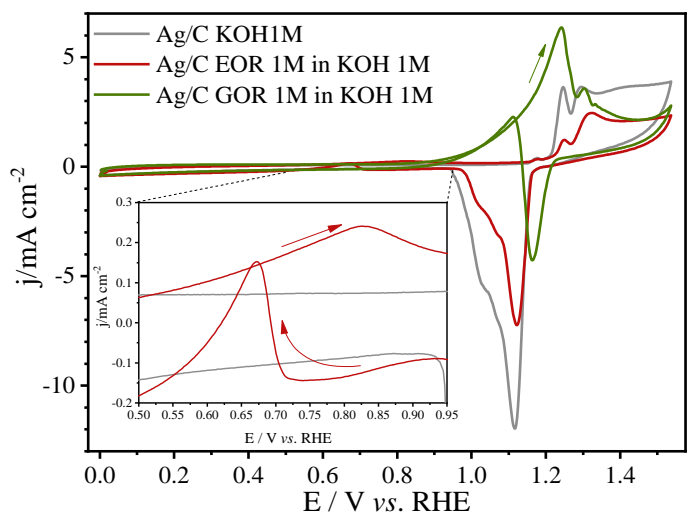

**Figure 7.** Cyclic voltammograms for Ag/C in KOH (1.0 mol L$^{-1}$, black), in ethanol (1.0 mol L$^{-1}$, red), and glycerol (1.0 mol L$^{-1}$, green), both in KOH (1.0 mol L$^{-1}$), ν = 20 mV s$^{-1}$.

*Figures 8a and 8b present the chronoamperometric profiles for ethanol oxidation at 0.70 V and glycerol oxidation at 0.80 V, respectively*. It presents data for the Pd/C ref, Ag/C, AgPd/C, AgAu/C, and AgPdAu/C electrocatalysts. In the first few seconds of chronoamperometry (CA), a sudden drop in current occurs due to the double-layer charging, and even peaks of system instability can be observed under the experimental conditions. After this brief time interval, the system stabilizes.[21]

For the EOR and GOR chronoamperometry studies, the measurements show a slight drop in current density, likely due to the lack of fuel restoration and/or electrode degradation or poisoning, resulting from the formation and adsorption of intermediate species generated during the electrooxidation of alcohols within the measured time interval.[23] According to Figure 8a, this drop is less pronounced for the AgPd/C compared to AgPdAu/C. The AgPd/C electrocatalyst stands out because, even with a reduction of approximately 75% in Pd loading compared to the commercial Pd/C ref., the current at the end of the experiment was very similar, possibly due to the contribution of Ag. Ag aids in the activation of ethanol by interacting with hydroxyl species (OH⁻), thereby favoring the removal of intermediates and improving the reaction efficiency, while also reducing the amount of palladium (Pd) required, making the catalyst more economical without compromising performance.[48] As expected from the CV experiments, the Ag/C and AgAu/C electrocatalysts resulted in currents close to zero, indicating no activity for EOR.

In GOR, the presence of metal hydroxides (AgOH, Au(OH)$_3$) and oxides (Ag$_2$O, Au$_2$O$_3$) improves the adsorption and activation of OH⁻ species, favoring the removal of hydrogen from the molecule and accelerating the initial stages of the reaction, which corroborates the results for the AgPdAg/C material, which resulted in higher current densities values, obtained from CA, than other electrocatalysts.[21,25,47,49]

The increasing order of current density from the EOR chronoamperometry is: AgPdAu/C (9 mA cm$^{-2}$) < AgPd/C (20 mA cm$^{-2}$) < Pd/C ref. (23 mA cm$^{-2}$). The AgPd/C and AgPdAu/C electrocatalysts also show promise in GOR studies. The ternary AgPdAu/C proved to be active and stable, resulting in the highest current density, even higher than that obtained with the commercial Pd/C ref., despite a 75% reduction in Pd loading, which can be explained by the presence and contribution of Au in the material composition, as Au nanoparticles can prevent the dissolution of Ag or Pd through electronic and geometric stabilization effects, in addition to modifying the binding energies of intermediates, affecting catalytic activity and inducing functional restructuring, in which the surface adapts to optimize the reaction for GOR in alkaline medium.[21,50] The increasing order of current density from the GOR chronoamperometry is: Ag/C (0.04 mA cm$^{-2}$) < AgAu/C (0.3 mA cm$^{-2}$) < AgPd/C (5 mA cm$^{-2}$) < Pd/C ref. (7 mA cm$^{-2}$) < AgPdAu/C (9 mA cm$^{-2}$).

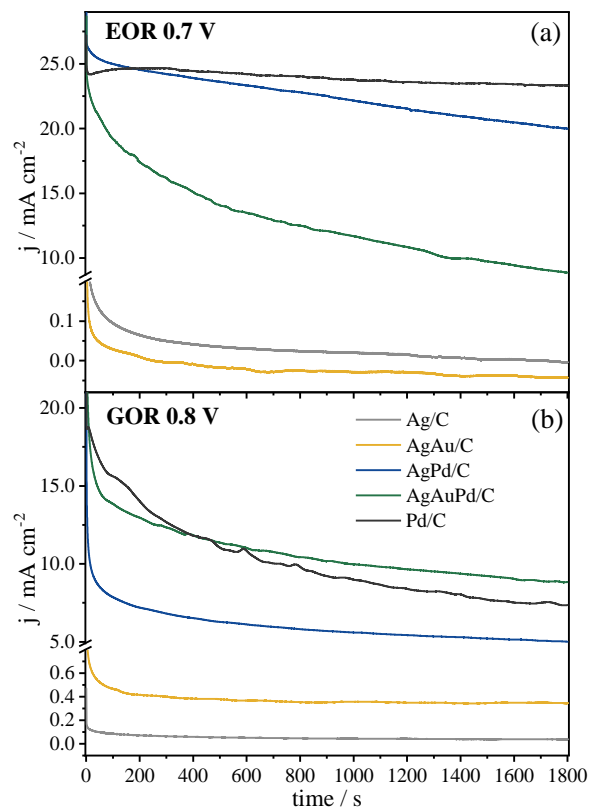

**Figure 8.** Chronoamperometric curves for Pd/C ref., Ag/C; AgPd/C; AgAu/C, and AgPdAu/C in **(a)** ethanol (1.0 mol L$^{-1}$) at 0.7 V, and **(b)** glycerol (1.0 mol L$^{-1}$) at 0.8 V, both in KOH (1.0 mol L$^{-1}$).

A comparison among the current density activities from CV, chronoamperometry curves, and the onset potentials for each electrode can be seen in Tables 3 and 4.

**Table 3**. Current density peak values obtained from cyclic voltammetry (CV), $E_{Onset}$, and current density from chronoamperometry (CA), for EOR and GOR.

|  | EOR | | | GOR | | |
|---|---|---|---|---|---|---|
|  | CV $j$ / mA cm$^{-2}$ | $E_{Onset}$ / mV vs. RHE | CA / mA cm$^{-2}$ | CV $j$ / mA cm$^{-2}$ | $E_{Onset}$ / mV vs. RHE | CA / mA cm$^{-2}$ |
| **Pd/C** | 32 | 326 | 23 | 32 | 410 | 7 |
| **Ag/C** | (-) | (-) | (-) | (-) | (-) | 0.04 |
| **AgPd/C** | 26 | 266 | 20 | 23 | 360 | 5 |
| **AgAu/C** | (-) | (-) | (-) | 28 | 610 | 0.3 |
| **AgPdAu/C** | 29 | 276 | 9 | 29 | 340 | 9 |

**Table 4.** Onset potentials, CA retention, and $J_f/J_b$ comparison for all electrodes studied.

| | | Ethanol | | |
|---|---|---|---|---|
| **Electrode** | **$E_{onset}$ vs. RHE** | **Comparing with Pd/C** | **CA retention** | **$J_f/J_b$** |
| Pd/C | 326 mV | | 85.7 % | 0.99 |
| AgPd/C | 266 mV | ~60 mV less positive | 65.4 % | 1.30 |
| AgPdAu/C | 276 mV | ~50 mV less positive | 31.0 % | 1.13 |
| AgAu/C | Not active | | | |
| | | **Glycerol** | | |
| **Electrode** | **$E_{onset}$ vs. RHE** | **Comparing with Pd/C** | **CA retention** | **$J_f/J_b$** |
| Pd/C | 410 mV | | 28.5 % | 1.01 |
| AgPd/C | 360 mV | ~50 mV less positive | 22.6 % | 2.03 |
| AgPdAu/C | 340 mV | ~70 mV less positive | 30.8 % | 1.27 |
| AgAu/C | 610 mV | ~200 mV more positive | 7.9 % | 1.07 |

*Regarding AgAu/C, its lack of EOR activity but significant GOR activity can be explained based on the activation adsorption energy between the surface and the reactant. EOR requires efficient C–H and C–C activation, which on AgAu/C (without Pd) is limited; while GOR is more strongly assisted by oxophilic Ag/Au oxy-hydroxides, which favor multi-step dehydrogenation and formation of $C_1$–$C_3$ carboxylates, as confirmed by FTIR (formate, glycolate, carbonate bands in Figures 10–11).*

## 2.3. *In situ* FTIR Experiments for Alcohol Oxidation Reactions

Spectroelectrochemical experiments were conducted to investigate the electrooxidation mechanisms of ethanol and glycerol on supported nanoparticulate catalysts, specifically Pd/C, AgPd/C, and AgPdAu/C, using *in situ* FTIR. This technique provides direct evidence of the nature of the different species adsorbed and reaction products formed during electrocatalytic reactions.

### 2.3.1. *In situ* FTIR Study of Ethanol Electrooxidation

The collection spectra of *in situ* FTIR are displayed in Figure 9, which were collected over a potential range from 0.1 to 1.3 V with incremental steps of 0.1 V. According to established literature, ethanol oxidation can proceed via two distinct pathways: a complete oxidation pathway (C1) and a partial oxidation pathway (C2), summarized by the following reactions:

(C1): $CH_3CH_2OH + 12OH^- \rightarrow 2CO_2 + 9H_2O + 12e^-$

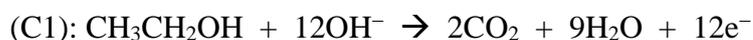

$$\text{(C2): } CH_3CH_2OH + 5OH^- \rightarrow CH_3COO^- + 4H_2O + 4e^-$$

Pathway C1, ethanol represents complete oxidation via C–C bond cleavage to $CO_2$, involving a 12-electron transfer. In contrast, pathway C2 represents partial oxidation, involving a 4-electron transfer and producing acetate ions. The greater the number of ethanol molecules converted to $CO_2$ via pathway C1, the higher the energy density obtained.[51,52]

Figure 9 presents the FTIR spectra for the Ag-based electrocatalysts. As shown in Figures 6-7, the Ag/C and AgAu/C materials showed negligible electroactivity towards EOR; thus, no FTIR spectra were recorded for these materials.

According to cyclic voltammetry data (Figure 6a), the EOR starts at approximately 0.35 V for the Pd-containing catalysts during the forward potential scan, coinciding with the emergence of two negative bands at approximately 1552 and 1414 $cm^{-1}$. These bands, assigned to the asymmetric and symmetric O–C–O stretching vibrations of the acetate ion ($CH_3COO^-$), respectively, strongly indicate acetate formation as the primary oxidation product.[53–55] In addition, a comparison between these bands' formation with the chronoamperometry curves (Figure 8) at a potential of 0.80 V, where the current density is relevant, indicates that the acetate production reaches a maximum value, as indicated in Figure 9. A weaker band at 1346 $cm^{-1}$ was also observed and attributed to acetate species in solution.[56,57] The dominance of acetate formation aligns with previous studies on Pd-based electrocatalysts in alkaline conditions, where acetate stability and strong surface adsorption are well documented.[53,58,59]

Additionally, a positive absorption bands were identified at approximately 1088 $cm^{-1}$. This band corresponds to the C–O stretching vibrations of ethanol molecules, indicating a decrease in ethanol concentration near the electrode interface, which confirms ethanol consumption during oxidation.[53,54,59]

Notably, no absorption band corresponding to $CO_2$ (expected around 2343 $cm^{-1}$) was detected on any of the catalysts tested under the specified experimental conditions. This absence strongly suggests a minimal or negligible C–C bond cleavage during ethanol oxidation, emphasizing the preferential route of ethanol oxidation through partial oxidation to acetate (C2 pathway). This observation is consistent with numerous prior reports in the literature where similar results have been documented for Pd, Au, Rh, and Pt electrocatalysts, as well as their alloys in alkaline medium, pointing out the pH-dependent difficulty of achieving complete ethanol oxidation to $CO_2$.[53,59] Additionally, there was no evidence of the $\nu(CO_{ad})$ signal related to linearly adsorbed CO species, indicating that CO is not a significant intermediate or poisoning species in the investigated potential region for these Ag-based electrocatalysts in alkaline medium. The lack of CO intermediates contrasts sharply with acidic media, where adsorbed CO species significantly hinder catalytic performance.[60]

The introduction of Au into Pd-based electrocatalysts, as demonstrated by Silva et al.[59], can induce beneficial electronic modifications, altering the d-band center of Pd and thereby modifying adsorption strengths for critical intermediates, such as acetyl and CO species. The authors demonstrated that such modifications could enhance electrocatalytic activity by facilitating acetyl oxidation, thereby promoting higher stability against poisoning. Similarly, Guo et al.[53] showed that alloying Pd with Ru increased the electronic vacancy of Pd's d-band, facilitating ethoxy formation but adversely affecting acetaldehyde oxidation. In this regard, the incorporation of Pd and Au in the Ag matrix in our materials likely induces analogous electronic effects, selectively stabilizing adsorbed acetyl intermediates and directing ethanol oxidation predominantly via the partial oxidation route to acetate.

Considering the acetyl intermediate formation, Yang et.al.[61] using the ATR-SEIRAS technique showed that ethanol oxidation on Pd electrodes in alkaline medium generally involves adsorbed acetyl species ($CH_3CO_{ads}$), identified explicitly by the presence of an absorption band approximately 1625 cm$^{-1}$. Unfortunately, in our experiments, this acetyl band overlapped with the interfacial water δ(HOH) deformation band located at approximately 1640 cm$^{-1}$, making precise identification challenging under our experimental conditions. In fact, EOR on palladium and gold electrodes shares similar mechanistic pathways with platinum.[61,18]

Moreover, it is noteworthy that although the formation of CO is negligible, the acetate formation, while stable and strongly adsorbed, can negatively impact catalyst stability over time, as seen in the chronoamperometric experiments (Figure 8). Guo et al.[53] identified catalyst poisoning caused by accumulated acetaldehyde dimers (formed via aldol condensation in alkaline medium) as a significant factor in performance degradation, particularly for Pd-based electrocatalysts. Although such dimers were not directly identified in our FTIR spectra, their potential formation under prolonged operation should be considered in future stability assessments.[57]

Collectively, the spectroscopic observations indicate that the ethanol oxidation mechanism on Pd/C, AgPd/C, and AgPdAu/C electrocatalysts in alkaline conditions predominantly proceeds via partial oxidation pathways. Like Pt, ethanol adsorbs forming adsorbed acetyl species ($CH_3CO_{ads}$) as an intermediate, which oxidizes directly to acetate ions, making acetate the primary stable product observed at relatively low potentials. In this way, ethanol is preferentially converted into stable acetate ions, consistent with the established mechanistic understanding of Pd-based catalysts in alkaline medium. The presence of Pd and Au within an Ag matrix, although influencing electronic properties and possibly affecting catalytic activity, did not notably alter the fundamental oxidation pathway or principal reaction intermediates identified in these spectroelectrochemical experiments. Nevertheless, the absence of adsorbed CO intermediates can be considered an advantage of working under alkaline environments with these catalysts, particularly when considering their use in fuel cells, for instance. Therefore, although alloying generally modifies the electronic properties of catalytic sites, the precise combination and distribution of metal components significantly influence the selectivity toward complete or partial ethanol oxidation.

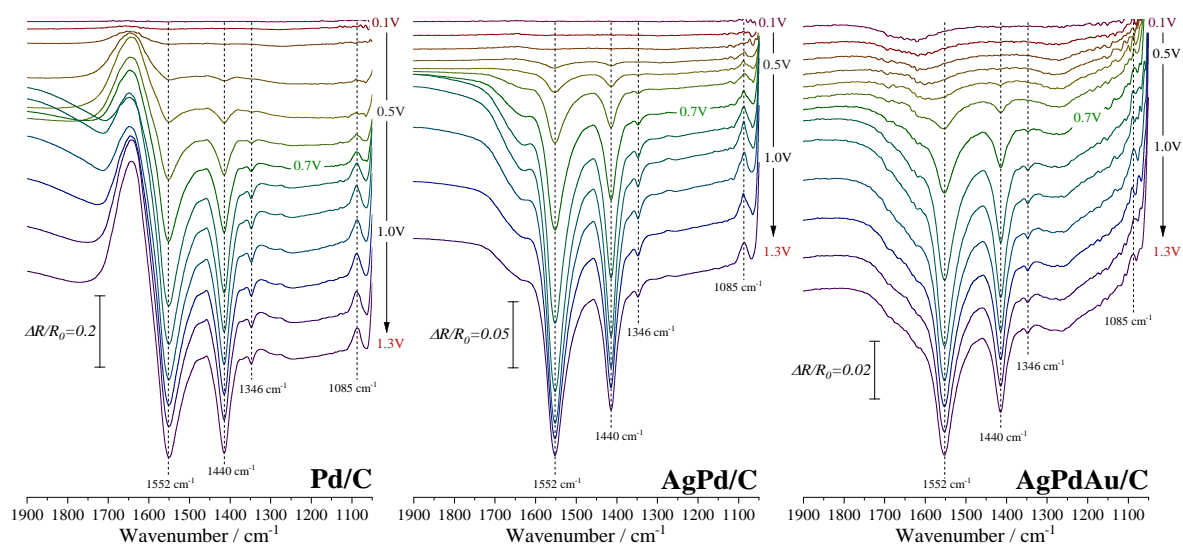

**Figure 9.** *In-situ* FTIR recorded in ethanol (1.0 mol L$^{-1}$) in KOH (1.0 mol L$^{-1}$) on: (a) Pd/C ref., (b) AgPd/C, and (c) AgPdAu/C. Potential range: E = 0.1 to 1.3 V, potential step: 0.1 V.

### 2.3.2. *In situ* FTIR Study of Glycerol Electrooxidation

The *in situ* FTIR spectra recorded during glycerol electrooxidation on Pd/C, Ag/C, AgPd/C, AgAu/C, and AgPdAu/C catalysts in 1.0 mol L$^{-1}$ glycerol + 1.0 mol L$^{-1}$ KOH are presented in Figure 10. Unlike simpler alcohols such as ethanol, glycerol possesses three hydroxyl groups, significantly increasing the complexity of its electrooxidation pathways and resulting in a diverse array of potential C1–C3 oxidation products.

For all the materials evaluated here, we observe that at low potentials (< 0.5 V), no distinct spectral bands are visible. Starting from approximately 0.5 V, several vibrational bands progressively appear across all examined electrocatalysts. Notably, a dominant broad negative band centered at approximately 1580 cm$^{-1}$ emerges and intensifies as the potential increases, and this band is further intensified at the potential at which the chronoamperometry curve was obtained (0.70 V), as shown in Figure 8. Concurrently, less intense bands are detected at approximately 1075, 1110, 1308, 1352, 1385, and 1410 cm$^{-1}$. Assigning these bands to single specific products is challenging due to the complexity of the reaction and overlapping characteristic vibrations of potential products; however, some selectivity related to the catalyst composition can be inferred. Table 5 summarizes the observed vibrational frequencies and their corresponding assignments.

Previous studies have consistently reported that incomplete oxidation dominates the glycerol electrooxidation reaction, leading to the formation of soluble intermediates, such as aldehydes, ketones, and carboxylates.[62–66] These species generate overlapping infrared absorption bands in the range from 1400 to 1580 cm$^{-1}$,[62,64,67] which can be more clearly visualized in Figure 11. Carboxylate species, in particular, exhibit two characteristic IR absorption bands: a strong asymmetrical stretching mode ($v_{as}$(O–C–O)) typically observed between 1550 – 1650 cm$^{-1}$ and a weaker symmetrical stretching mode ($v_s$(O–C–O)) observed between 1300 – 1420 cm$^{-1}$, with the precise frequencies dependent on the specific carboxylate formed.[67]

Considering this spectral overlap, the broad envelope observed between 1300–1420 cm$^{-1}$ range reveals a distinct band at around 1308 cm$^{-1}$, appearing at potentials above 0.75 V, particularly when Pd is present in the catalyst composition. This band can be assigned to the C–O stretching vibration characteristic of oxalate species.

The progressive oxidation of glycerol into smaller carboxylates through C-C bond cleavage, like glycolate (1075 cm$^{-1}$) and oxalate (1308 cm$^{-1}$), inherently leads to the formation of C1 oxidized species, such as formate, whose presence is indicated by the IR absorption band at 1352 cm$^{-1}$. However, the presence of this band does not strictly correlate with the addition of Au or Pd to the Ag matrix, since the vibrational mode of formate is also clearly visible in the FTIR spectra for Ag/C. This finding is consistent with previous work by Suzuki et al.[62] identified glycolate and formate as major oxidation products during exhaustive glycerol electrolysis on bulk Ag electrodes.

A careful examination of the symmetric stretching ν(OCO)s band reveals an additional spectral feature around 1370 cm$^{-1}$ for all catalysts investigated, though with varying intensities. This negative band is typically attributed to the symmetric stretching vibration of carbonate species formed via the complete oxidation of glycerol to $CO_2$. The detection of carbonate under strongly alkaline conditions further corroborates the occurrence of complete oxidation pathways during glycerol electrooxidation.

Another notable feature of the FTIR spectra is the broad band centered near 1640 cm$^{-1}$. While this band could potentially indicate the presence of adsorbed glycerol intermediates, the clear-cut assignment is not possible due to the bending mode of interfacial water within the thin-layer cavity of the spectroelectrochemical cell configuration. As previously discussed, this water band significantly interferes in this region, making a definitive assignment challenging.

Concurrently, a weak IR feature at approximately 1730 cm$^{-1}$ emerges, overlapping partially with the water bending vibration. This spectral band strongly suggests the formation of dihydroxyacetone (DHA), a primary oxidation product resulting from the selective oxidation of glycerol without C—C bond cleavage. Although the ν(C=O) feature at this frequency is relatively weak, its position closely corresponds to the characteristic stretching mode of DHA. Interestingly, this particular band is exclusively detectable for Pd/C in the present study, contrasting with previous findings by Coutanceau et al.[8,68] reported selective DHA formation only for specific Pd-alloy catalysts (Pd–Bi and Pd–Au).

Overall, the catalyst surface composition has a significant impact on the balance between the oxidation routes. Pd-rich surfaces (Pd/C and AgPd/C) exhibit greater selectivity toward partially oxidized carboxylates, notably oxalate. In contrast, Ag-rich surfaces (Ag/C and AgAu/C) preferentially catalyze deeper oxidation processes, yielding carbonate and smaller carboxylate species. The ternary catalyst AgPdAu/C exhibits intermediate behavior, underscoring the nuanced electronic and ensemble effects that arise from alloying.

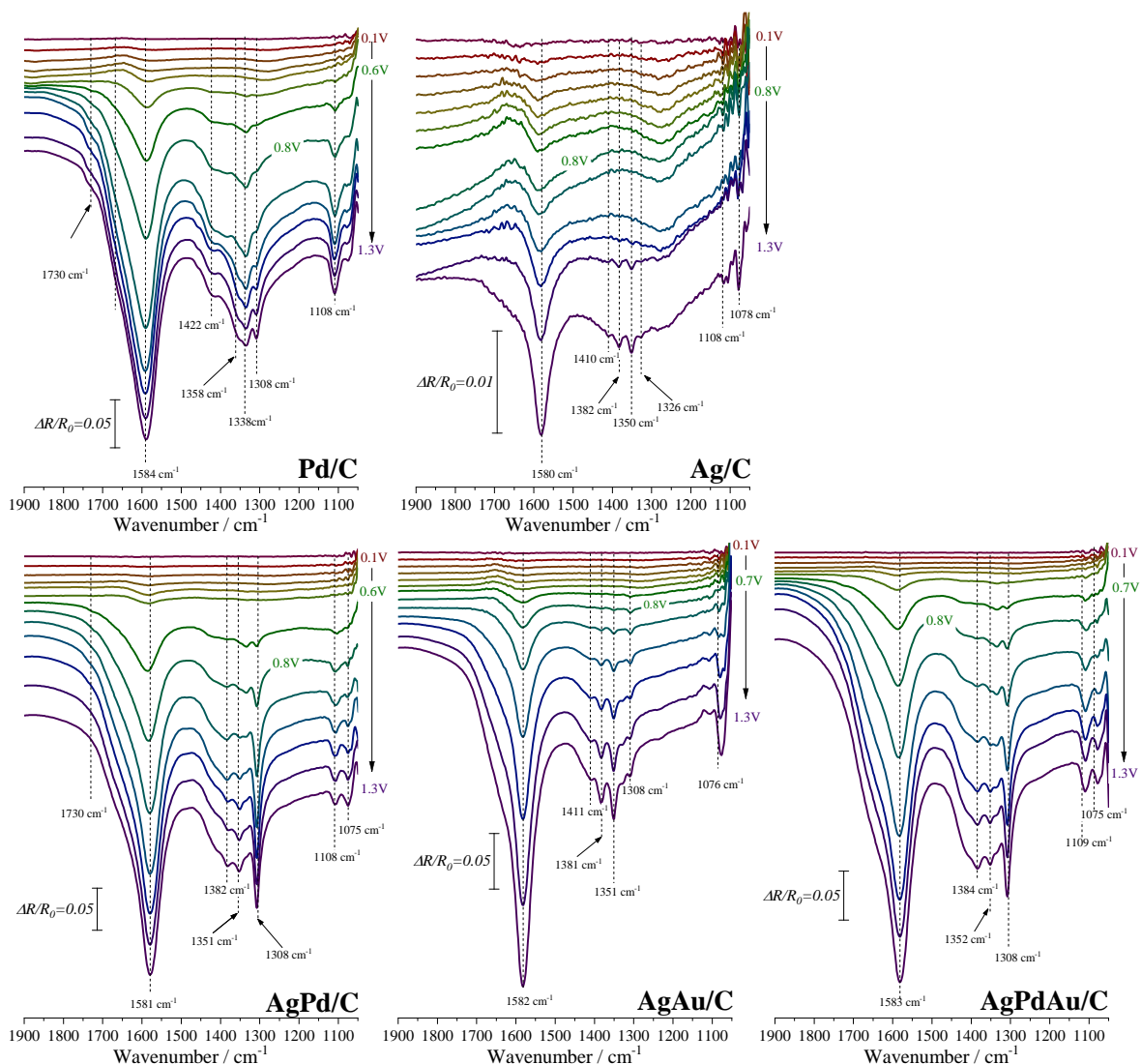

**Figure 10.** *In-situ* infrared spectra recorded in glycerol (1.0 mol L$^{-1}$) in KOH (1.0 mol L$^{-1}$) on Pd/C ref., Ag/C, AgPd/C, AgAu/C, and AgPdAu/C. Potential range: E = 0.1 to 1.3 V and potential step: 0.1 V.

**Table 5**. Vibrational frequencies and band assignments of the main spectral features are displayed in Figure 11.

| Ag (cm$^{-1}$) | Pd (cm$^{-1}$) | AgPd (cm$^{-1}$) | AgAu (cm$^{-1}$) | AgPdAu (cm$^{-1}$) | Assignment | Chemical Species | Reference |
|---|---|---|---|---|---|---|---|
| 1078 | 1075 | 1075 | 1075 | 1075 | | Glycolate | [69] |
| - | 1108 | 1108 | - | 1108 | $vs(C-O)$ | Alcohol | [62,67] |
| - | 1308 | 1308 | 1308 | 1308 | $vs(O-C-O)$ | Oxalate | [62] |
| 1326 | - | - | - | - | | | |
| - | 1335 | - | - | - | - | Dihydroxyacetone | [68,70] |
| 1350 | 1352 | 1352 | 1351 | 1352 | $vs(O-C-O)$ | Formate | [39,66,67] |
| 1382 | - | 1382 | 1381 | 1384 | $vs(O-C-O)$ | Carbonate/glycerate | [65] |
| - | 1415 | 1413 | 1411 | 1415 | $vs(O-C-O)$ | Glycolate | [71] |
| - | 1458 | - | - | - | | | |
| 1580 | 1584 | 1580 | 1583 | 1581 | $vas(O-C-O)$ | Carboxylates | [67] |
| - | 1650 | 1650 | 1650 | 1650 | | | |
| - | 1730 | - | - | - | $C=O$ | Carbonyl (ketones) | [67] |
| - | 2300 | 2300 | - | - | $O=C=O$ | $CO_2$ | [67] |

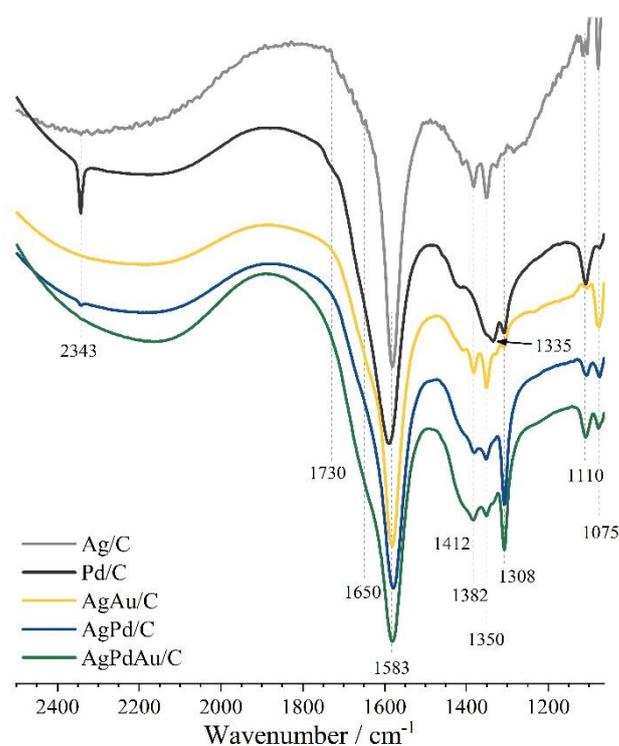

**Figure 11.** *In-situ* infrared spectra recorded in glycerol (1.0 mol L$^{-1}$) in KOH (1.0 mol L$^{-1}$) on Pd/C ref., Ag/C, AgPd/C, AgAu/C, and AgPdAu/C. Potential applied: E = 0.1 to 1.3 V and potential step: 0.1 V.

*According to FTIR results, it is possible to infer the main pathways observed for the glycerol reaction. Then, a schematic representation of the main glycerol oxidation pathways and representative alkaline reactions leading to $C_1$—$C_2$ products (glycerate, glycolate, oxalate, formate, and carbonate)* [72] *is presented through Figure 12.*

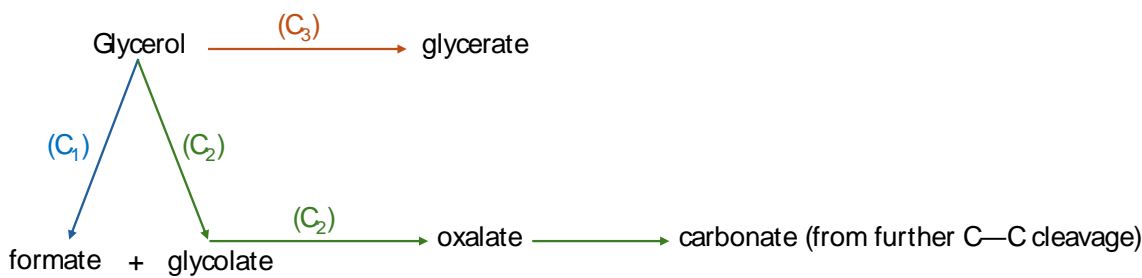

**Figure 12**. Schematic representation of the main glycerol oxidation pathways.

## 3. Conclusion

*This work demonstrates that Ag-based mono-, bi- and trimetallic catalysts with drastically reduced Pd loading exhibit competitive electrocatalytic activity toward ethanol and glycerol oxidation in alkaline media. AgPd/C and AgPdAu/C, containing only 5 wt% Pd, display onset potentials 50–70 mV lower than commercial Pd/C and exhibit enhanced stability and resistance to poisoning. FTIR results confirm acetate as the main EOR product and carboxylate/carbonate species as dominant GOR products. The oxophilic nature of Ag, combined with the electronic tuning induced by Pd and Au, governs activity and durability. These findings highlight Ag-rich multimetallic catalysts as promising low-cost alternatives for alkaline alcohol electrooxidation.*

## 4. Experimental Section

*Synthesis of Nanoparticles (NPs):* To obtain the metallic nanoparticles, the chemical reduction method[9] using sodium borohydride was employed, with Vulcan XC-72 carbon from Cabot Corporation as the support. The synthesized electrocatalysts were: AgPd/C ($Ag_{40\%}Pd_{5\%}/C_{55\%}$), AgAu/C ($Ag_{40\%}Au_{5\%}/C_{55\%}$), AgPdAu/C ($Ag_{35\%}Pd_{5\%}Au_{5\%}/C_{55\%}$), and Ag/C ($Ag_{45\%}/C_{55\%}$) – with percentages given by mass ratio. The metal precursors were silver nitrate ($AgNO_3$), palladium nitrate ($Pd(NO_3)_2$), and chloroauric ($HAuCl_4$). Initially, the metal precursors were added to a beaker and dissolved in 20 mL of a 1:1 (v/v) mixture of $H_2O$ and isopropanol. After dissolution, the mixture was stirred for 15 min under mechanical agitation, followed by 10 min of ultrasound agitation. An excess solution of sodium borohydride ($NaBH_4$) was added at a 1:20 molar ratio between the metal and $NaBH_4$. Subsequently, the mixture was stirred for 30 min under mechanical agitation, Vulcan XC-72 carbon was added, and the mixture was maintained under mechanical agitation for an additional 2 h and 30 min. The material was then washed with a 1:1 (v/v) mixture of $H_2O$ and ethanol and dried in an oven at 90 °C for 24 h. *The catalyst compositions were selected to maintain a fixed total metal loading of 45 wt% while minimizing high-cost metal content. Only 5 wt% Pd (and 5 wt% Au when present) was used. This strategy enables direct comparison with commercial Pd/C while significantly reducing noble-metal content.*

*Preparation of NPs/C Catalyst Ink:* The NPs/C catalyst ink was prepared by dispersing approximately 8 mg in 1 mL of a mixture of ultrapure water and isopropanol (80:20 v/v, respectively). The dispersion was sonicated for 20 min in an ice bath. After sonication, 20 µL aliquots of the ink were deposited onto the glassy carbon (GC) electrode (0.196 cm$^2$) using the drop-casting technique. The electrode was dried under a lamp, and after drying, the same volume of a 5% Nafion solution in ultrapure water (1:100 v/v, respectively) was added and dried again under a lamp.

*Physical-chemical Characterization:* Powder XRD patterns were recorded on a Bruker AXS D8-Advanced X-ray diffractometer with CuKα radiation, λ =1.5418 Å). The metallic nanoparticles supported on carbon were characterized by high-resolution transmission electron microscopy (HRTEM) in a JEOL/JEM-3010 microscope operating at 300 kV. The metal loading was first determined by differential thermal analysis (DTA) and thermogravimetric analysis (TGA). Then, an estimation of the elementary particle composition was performed using energy-dispersive X-ray analysis (EDS) and quantitatively by inductively coupled plasma mass spectrometry (ICP-MS, Agilent 7900, Hachioji, Japan), operated with high-purity argon (99.999%), White Martins.

*Electrochemical measurements:* The Autolab 302N potentiostat was used for electrochemical characterization and studies on alcohol oxidation. In the electrochemical cell, 1.0 mol L$^{-1}$ KOH was used as the electrolyte, and the experiments were performed at room temperature in a deoxygenated medium by purging argon into the electrolyte for approximately 10 min before each measurement. For the alcohol oxidation reaction, the same experimental conditions were used, and 1.0 mol L$^{-1}$ of each fuel was added to the electrochemical cell. *For this, three-electrode glass cell configuration (glassy carbon modified by nanoparticles as working electrode, Pt counter electrode, Ag/AgCl reference electrode) was used.* The Ag/AgCl reference electrode was used, along with a Pt electrode and a GC working electrode. The pH of the electrolyte solutions was measured and used to convert the applied potentials *versus* (*vs*.) the Ag/AgCl reference electrode (E$_{applied}$) to the reversible hydrogen electrode

(RHE) scale (E$_{RHE}$), using the following equation, where E$_{Ag/AgCl}$ is the potential of the Ag/AgCl reference electrode.[6]

$$E_{RHE} = E_{applied} + E_{Ag/AgCl} + 0.0591 \times pH$$

*In situ FTIR:* The *in situ* FTIR experiments were performed using a VERTEX 70V Bruker spectrometer equipped with an LN-MCT Mid detector cooled with liquid nitrogen. *A semi-spherical CaF$_2$ was used as the infrared window, and the working electrode was pressed against this window, so that a thin electrolyte layer (thickness of 1 – 10 μm) is trapped between them. The cell was operated under static conditions (no electrolyte flow), following standard thin-layer IRRAS protocols reported for alcohol electrooxidation* [73]. The reference spectrum (R$_0$) was collected at 0.05 V in the same sample solution containing ethanol (1.0 mol L$^{-1}$) for EOR, and glycerol (1.0 mol L$^{-1}$) for GOR + KOH (1.0 mol L$^{-1}$). *In situ* FTIR data were collected simultaneously during successive potential steps of 0.10 V up to 1.30 V. The spectra were computed from the average of 256 interferograms, and the spectral resolution was set to 4 cm$^{-1}$ and the *scanner velocity 160 kHz. The IR beam hits the window–electrolyte–electrode stack at near-grazing incidence, angles between 70–80°.*

## Supporting InformationSupporting

Information is available from the Wiley Online Library or from the author.

## Acknowledgements

T.C.G. and G.T.F. acknowledge financial support from FAPESP (grants #2023/04297-0 and #2019/22183-6). G.T.F. acknowledges financial support from CNPq (grant #3134455/2021-0). G.T.F. acknowledges the support of the RCGI - Research Center for Gas Innovation, hosted by the University of São Paulo (USP) and sponsored by FAPESP (2020/15230-5) and Shell Brazil (CW410367), and the strategic importance of the support given by ANP (Brazil's National Oil, Natural Gas, and Biofuels Agency) through the R&D levy regulation. V.D.C. acknowledges financial support from CNPq (grant #312788/2022-3). C.A.A. acknowledges financial support from FAPESP (#2018/10292-2 and #2022/11345-8). M.C.S. acknowledges FAPESP (#2022/15252-4).

## Conflict of Interest

The authors declare no conflict of interest.

**Author Contributions**

Author contributions are defined based on the CRediT and listed alphabetically. Conceptualization was performed by V.D.C., G.T.F., and T.C.G. Data curation was provided by C.A.A., T.C.G., and V.D.C. Formal analysis was provided by V.D.C. and T.C.G. Funding acquisition was provided by G.T.F. Investigation was performed by B.L.B., C.N.L., C.A.A., H.S.N.L., T.C.G., and V.D.C. Methodology was performed by T.C.G., and V.D.C. Project administration was provided by T.C.G. and V.D.C. Resources were provided by G.T.F. and V.D.C. Software was provided by C.A.A., V.D.C., and T.C.G. Supervision was done by G.T.F., M.C.S., and V.D.C. Validation was provided by G.T.F., T.C.G., and V.D.C. Visualization was done by V.D.C. and T.C.G. Writing – original draft was done by C.A.A., G.T.F., M.C.S, and V.D.C., T.C.G. Writing, review, and editing were done by all.

**Data Availability Statement**

The data that support the findings of this study are available from the corresponding author upon reasonable request.

**Keywords**

ethanol and glycerol electro-oxidation reaction, silver electrocatalysts, low content of high-cost metals, *in situ* FTIR, electrochemical techniques